\documentclass[journal]{IEEEtran}
\usepackage{algpseudocode}                                 
\usepackage{algorithm}
\usepackage{graphicx}                                      
\usepackage{amsmath}
\usepackage{amssymb}
\usepackage{amsfonts}
\usepackage{booktabs}
\usepackage{multirow}
\usepackage{authblk}
\usepackage[subrefformat=parens,farskip=0pt,justification=centering]{subfig}
\usepackage{color}
\usepackage{cite}                                          
\usepackage{comment}                                       
\usepackage{soul}                                          
\soulregister\cite7
\soulregister\ref7
\soulregister\pageref7
\usepackage{amsthm}
\usepackage{etoolbox}                                      
\usepackage{url}
\usepackage{nth}                                           
\usepackage{bm}                                            
\usepackage{courier}
\usepackage{balance}
\usepackage{threeparttable}
\usepackage[bookmarks=false]{hyperref}                     %
\hypersetup{
    colorlinks = true,
    citecolor  = blue,
    linkcolor  = blue,
    urlcolor   = blue,
}
\usepackage{filecontents}                                  
\usepackage{pgfplots}
\usepackage{pgfplotstable}
\pgfplotsset{compat=newest}
\usepackage{caption}
\algrenewcommand\textproc{\texttt}

\makeatletter
\let\OldStatex\Statex
\renewcommand{\Statex}[1][3]{%
  \setlength\@tempdima{\algorithmicindent}%
  \OldStatex\hskip\dimexpr#1\@tempdima\relax
}
\makeatother

\RequirePackage[normalem]{ulem} 
\RequirePackage{color}\definecolor{RED}{rgb}{1,0,0}\definecolor{BLUE}{rgb}{0,0,1} 

\begin{document}

\title{\Large\bf
Spintronics based Stochastic Computing for Efficient Bayesian Inference System
}

\author[1]{\normalsize {Xiaotao Jia}}
\author[2]{\normalsize {Jianlei Yang}\thanks{This work was supported in part by the National Natural Science Foundation of China (Grand No. 61602022) and the 111 Talent Program B16001.}}
\author[1]{\normalsize {Zhaohao Wang}}
\author[3]{\normalsize {Yiran Chen}}
\author[3]{\normalsize {Hai (Helen) Li}}
\author[1]{\normalsize {Weisheng Zhao}}
\affil[1]{\normalsize{Fert Beijing Research Institute, BDBC, School of Electronic and Information Engineering, }}
\affil[2]{\normalsize{Fert Beijing Research Institute, BDBC, School of Computer Science and Engineering, }}
\affil[ ]{\normalsize{Beihang University, Beijing, 100191, China}}
\affil[3]{\normalsize{Department of Electrical and Computer Engineering, Duke University, Durham, NC 27708, U.S.A.}}
\affil[ ]{\normalsize{Email: jianlei@buaa.edu.cn}}
\renewcommand\Authfont{\small}
\renewcommand\Affilfont{\small}

\maketitle
\thispagestyle{empty} 

{\small\bf Abstract---
Bayesian inference is an effective approach for solving statistical learning problems especially with uncertainty and incompleteness.
However, inference efficiencies are physically limited by the bottlenecks of conventional computing platforms.
In this paper, an emerging Bayesian inference system is proposed by exploiting spintronics based stochastic computing.
A stochastic bitstream generator is realized as the kernel components by leveraging the inherent randomness of spintronics devices.
The proposed system is evaluated by typical applications of data fusion and Bayesian belief networks. Simulation results indicate that the proposed approach could achieve significant improvement on inference efficiencies in terms of power consumption and inference speed.
}


\section{Introduction} \label{Section:Intro}

The rise of deep learning has greatly promoted the development of artificial intelligence, however, most modern deep learning models face several difficulties such as the requirement of large scale training data and overfitting problem during learning. Furthermore, they can neither represent the uncertainty and incompleteness of the world nor take advantages of well-studied experience and theories.
In order to overcome these limitations, some researches trend to utilize Bayesian inference or combine Bayesian approaches with deep learning. Bayesian inference provides a powerful approach for information fusion, reasoning and decision making that has established it as the key tool for data-efficient learning, uncertainty quantification and robust model composition. It is widely used in applications of artificial intelligence and expert systems, such as multisensor fusion~\cite{pinheiro2004Bayesian} and Bayesian belief network~\cite{cruz2007diagnosis}. Recent years, Bayesian approaches attract the attention of neural network researches. Several studies (such as~\cite{gal2017deep}) have been proposed to combine advances in Bayesian approaches into neural network learning.

Bayes' theorem is the theoretical foundation of Bayesian inference and the key operation is probabilistic computing. The implementation of probabilistic algorithms on floating-point architecture has some disadvantages such as inefficiency in terms of power consumption, computing speed and memory usage and the inability to exploit parallelism of the Bayesian inference~\cite{thakur2016Bayesian}. Further, as the scaling of feature size of transistor, physical phenomena, such as low noise margin, low supply voltage, manufacturing process variations and soft errors, makes traditional integrated circuits much error-prone~\cite{alaghi2013survey}. Consequently, unconventional computing method - stochastic computing (SC), that directly addresses these issues has attracted much attention. Enable very low-cost implementations of arithmetic operations using standard logic elements and high degree of error tolerance are two main attractive features of stochastic computing~\cite{alaghi2013survey}.

\begin{figure}[tb!]
    \centering
    \includegraphics{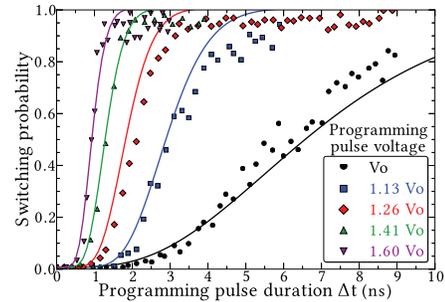}
    \caption{Experimental measurements of the
switching probability with respect to the duration of the applied programming
pulse, for different programming voltages.\cite{vincent2015analytical}}
    \label{fig:stochswitch}
\end{figure}

\begin{figure}[tb!]
    \centering
    \subfloat[Traditional SBG]{\includegraphics[scale=0.7]{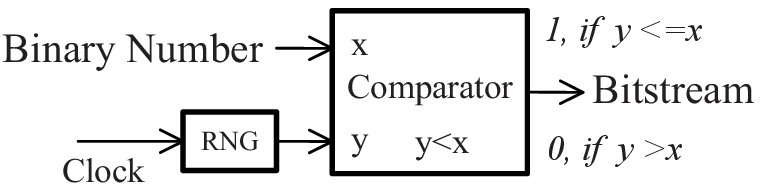}\label{fig:sbg:a}}
    \hspace{+10cm}
    \subfloat[$o_1 \approx p_1*p_2$]{\includegraphics[scale=0.65]{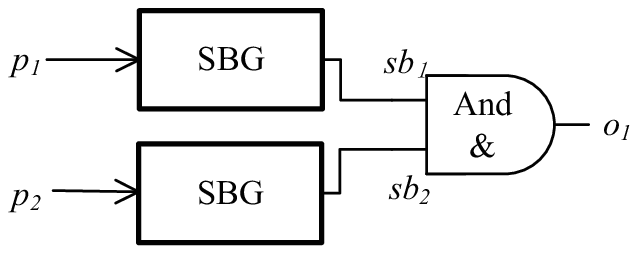}\label{fig:sbg:b}}
    \hspace{.1in}
    \subfloat[$o_2 \approx p_1*p_3+p_2*(1-p_3)$]{\includegraphics[scale=0.65]{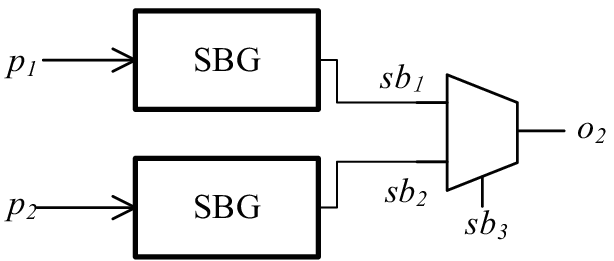}\label{fig:sbg:c}}
    \caption{(a) SBG circuit; (b) Multiplication operation using \texttt{AND} gate; (c) scaled addition operation using multiplexer (\texttt{MUX}).}
    \label{fig:sbg}
\end{figure}

The separation of processing units and memories remains a fundamental principle of von Neumann architecture computers even though there are many efforts towards increasing parallelism~\cite{grollier2016spintronic}. In order to improve Bayesian inference efficiency, several different specific hardware or circuits have been proposed such as FPGA~\cite{lin2010high} and analog circuits~\cite{mroszczyk2014accuracy}. Even though these works make an improvement on inference efficiency, there are still some shortcomings with the consideration of stochastic computing. Stochastic computing is executed using stochastic bitstreams. In most previous works, stochastic bitstreams (SB) are generated utilizing pseudo-random number generators (RNG) and comparators as shown in Fig.~\subref*{fig:sbg:a}. Unfortunately, generating (pseudo-)random bits is fairly costly. Therefore, the gate-level advantage of stochastic computing is typically lost. Towards to resolve these shortcomings, emerging nanometer-scale devices such as spintronics are considered as the major breakthroughs. In particular, magnetic tunnel junctions (MTJ) are well suited for bitstream generation because of its attractive feature such as non-volatile, low power and stochastic (Fig.~\ref{fig:stochswitch}). Several strategies have been proposed to generate stochastic bitstreams with spintronic devices~\cite{de2015stochastic,wang2016novel,wang2017hybrid}. However, shortcomings still exist in terms of power, area or speed. And none of them explain how to incorporate the stochastic bitstream generator with real world applications.

In this paper, a Bayesian inference system with less power consumption and high inference speed is built by stochastic computing based on spintronic devices and applied to traditional Bayesian inference applications.
The main contributions of this work are listed as follows:
\begin{itemize}
	\item A complete scheme of MTJ based stochastic bitstream generator (SBG) is proposed.
	Simulation results indicate that the stochastic bitstreams generated by the proposed SBG are with high accuracy and low correlation.
	\item Two efficient Bayesian inference systems are proposed utilizing the SBG and applied to data fusion and Bayesian belief network. Simulation results show that both two applications could achieve reasonable results with less energy, higher speed.
\end{itemize}

The remainder of this paper is organized as follows.
Section~\ref{Section:Preliminaries} states some preliminaries and related works.
The diagram of Bayesian inference system is illustrated in Section~\ref{Section:Diagram}.
Section~\ref{Section:sbg} describes details of SBG.
Bayesian inference systems for two real world applications are proposed in Section~\ref{Section:application}.
Finally, conclusion is given in Section~\ref{Section:Conclusion}.

\section {Background} \label{Section:Preliminaries}
\subsection{Stochastic Computing}
SC was first introduced in the 1960's by von Neumann~\cite{von1956probabilistic}. The basic idea of SC is that a number is presented by the ratio of `1' in a SB and arithmetic operations are implemented using simple logic gate(s) as shown in Fig.~\subref*{fig:sbg:b}\subref{fig:sbg:c}. It is worth to note that SBs which are highly correlated are not as expected, because higher correlation would lead to lower computing precision. In order to meet the requirement of sufficient random and uncorrelated, pioneer researcher proposed several SBG models such as linear feedback shift registers (LFSRs)~\cite{jeavons1994generating}, weighted binary SNG~\cite{gupta1988binary}. However, these CMOS based SBGs consume too much energy and area.

\vspace{-2mm}
\subsection {MTJ Basics}

The core part of the MTJ is a sandwich structure consisting of two ferromagnetic (FM) layers sandwiched with a tunneling barrier.
One FM  layer is called as reference layer with fixed magnetization direction. The other FM  layer is called as free layer whose magnetization direction could be parallel (\texttt{P}) or antiparallel (\texttt{AP}) with that of reference layer.
Because of the tunnel magnetoresistance effect, the nanopillar resistance depends on the relative orientation (\texttt{P} or \texttt{AP}) of the magnetization directions of the two FM layers.
An applied field can switch the free layer between the two directions.
The stochastic behavior of MTJ switching has been revealed by~\cite{devolder2008single}, which results from the unavoidable thermal fluctuations of magnetization~\cite{marins2012precessional}.
The stochastic switching is very suitable for generating stochastic bitstreams.

Recently, the work in~\cite{de2015stochastic} proposes an SBG based on MTJ.
But the circuit is too simple, and its implementation may be incomplete.
Furthermore, it does not consider the correlation of different SBs which may result in inaccuracy of SC.
A novel computing system using stochastic logic built by voltage-controlled MTJs (VC-MTJs) is proposed in~\cite{wang2017hybrid}.
This system consumes less energy and circuit area compared with LFSR circuits.
But in this system, the bit generation still involves too many MTJs and transistors.
Bitstream correlation is considered in this paper, but the proposed shuffle operation could not remove the relevance essentially and arithmetic operations between them maybe result in an unexpected number.
For example, a bitstream ${sb}_1(`10101010')$ presenting $0.5$ will be turned into ${sb}_2(`01010101')$ with the proposed shuffle operation in~\cite{wang2017hybrid}.
However, the result of ${sb}_1 \& {sb}_2$ will be $0$ rather than $0.25$.

\section {Diagram of Bayesian Inference system} \label{Section:Diagram}
\begin{figure}[tb!]
    \centering
    \includegraphics[scale=0.85]{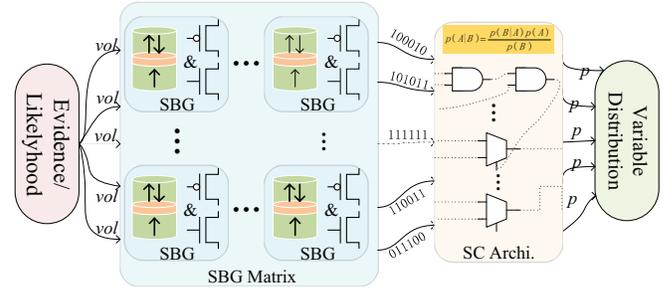}
    \caption{Diagram of the proposed Bayesian inference system.}
    \label{fig:diagram}
\end{figure}
Fig.~\ref{fig:diagram} describes the diagram of the proposed Bayesian inference system (BIS).
The input of BIS is a series of bias voltages corresponding to evidence or likelihood.
These evidences or likelihood may come from sensors in robot, autonomous, etc., also may come from clearly fact such as the X-ray results in Bayesian belief net for cancer diagnosis.
SBG matrix within light blue rectangle and SC architecture within light yellow rectangle are two key components of BIS.
SBG matrix is utilized to generate SBs based on input voltages.
Its scale is related with evidence count and variable relations.
Each SBG is a hybrid MTJ/CMOS circuit yielding SB with fast speed, low power and high accuracy.
Details of SBG are described in Section~\ref{Section:sbg}.
SC architecture is constructed by simple logic gates such as \texttt{AND} gate or multiplexer (\texttt{MUX}) and takes SBs as inputs.
The goal of SC architecture is to implement Bayesian inference utilizing SBs and SC theory on the basis of Bayes' Rule.
In this architecture, stochastic computing is achieved by a novel arrangement of \texttt{AND} gates and \texttt{MUX}s and the interconnections between them.
Usually, different applications are solved by different inference algorithms, thus, require different computing architectures which could be found in Section~\ref{Section:application}.
Finally, inference results are presented by the format of random variable distribution which could provide guidance for decision making.

\section {MTJ based stochastic Bitstream generator}
\label{Section:sbg}

Accuracy of Bayesian inference is mainly determined by the quality of bitstreams.
A ``Good'' bitstream should accurately represent a given probability number and also have low correlation with other bitstreams.
In this section, we introduce an SBG utilizing stochastic switching behavior of MTJ and then exhibit the simulation results.

\vspace{-2mm}
\subsection {Schematic of SBG}
\begin{figure}[tb!]
    \centering
    \includegraphics[scale=1]{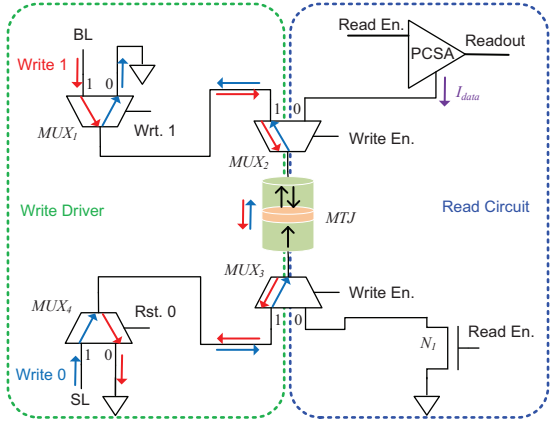}
    \caption{Schematic of SBG circuit.}
    \label{fig:sbg_schematic}
\end{figure}
In the proposed system, every bitstream is constructed based on the state of MTJ.
If MTJ is with high resistance \textit{i.e.} `\texttt{AP}' state,  `0' will be added to the bitstream; otherwise, `1' will be added.
Generally, the state of MTJs could be easily detected by CMOS sense amplifiers.

The circuit diagram of proposed SBG is illustrated as Fig.~\ref{fig:sbg_schematic} which is composed by CMOS transistors and MTJs.
Both write and read operations could be achieved with this circuit.
Bit-line (BL) and source-line (SL) are driven by two different voltage sources.
\texttt{MUX}$_2$ and \texttt{MUX}$_3$ are used to control either read current or write current would go through the MTJ.
During the write operation, signal `Write En' is at high level, thus terminal `1' of \texttt{MUX}$_2$ and \texttt{MUX}$_3$ are ON.
The write operation consists of two phases: resetting MTJ state to `\texttt{AP}' state and switching the MTJ state from `\texttt{AP}' to `\texttt{P}'.
In the first phase, terminal `0' of \texttt{MUX}$_1$ and terminal `1' of \texttt{MUX}$_4$ are ON because signal `Wrt. 1' is at low level and signal `Rst. 0' is at high level.
Current flows through the MTJ from bottom to top as the blue arrow shows.
In this phase, bias voltage and duration time are set to guarantee that the state of MTJ switches to `\texttt{AP}' state at 100\% probability.
In the second phase of write operation, terminal `1' of \texttt{MUX}$_1$ and terminal `0' of \texttt{MUX}$_1$ are ON because signal `Wrt. 1' is at high level and signal `Rst. 0' is at low level.
Current flows through the MTJ from top to bottom as the red arrow shows.
In this phase, bias voltage and duration time are set based on the wanted probability value.
During the read operation, terminal `0' of \texttt{MUX}$_2$ and \texttt{MUX}$_3$ and transistor \texttt{N}$_1$ are ON.
A Pre-Charge Sense Amplifier (PCSA)~\cite{zhao2009high} is used to read the state of MTJ.

\begin{figure}[tb!]
    \centering
    \includegraphics[scale=0.4]{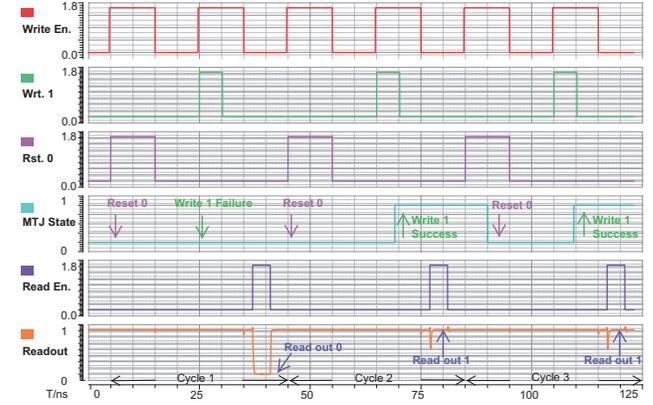}
    \caption{Simulation results of SBG circuit.}
    \label{fig:wave}
\end{figure}

Three-cycle Cadence simulated waveform is illustrated in Fig.~\ref{fig:wave}. Each cycle consists of three operations of resetting 0, writing 1 and reading MTJ state.
In each cycle, the MTJ state is first reset to be `\texttt{AP}' state during which `Write En' and `Rst. 1' are high and `Wrt. 1' is low.
Current goes through the MTJ from bottom to up as the blue arrow shows in Fig. \ref{fig:sbg_schematic}.
Then comes writing 1 stage during which `Rst. 0' is low and `Wrt. 1' is high.
Current goes through the MTJ from up to bottom as the red arrow shows in Fig. \ref{fig:sbg_schematic}.
In this stage, the state of MTJ may or may not switch from `\texttt{AP}' to `\texttt{P}'.
Then comes the reading stage during which `Read En.' becomes high and `Write En.' becomes low.
In this stage, the state of MTJ is read out by PCSA circuit as the last wave shows.
In the given example, writing 1 operation fails in the first cycle and successes in the following two cycles.
Thus, bitstream is generated as `011'.

\vspace{-2mm}
\subsection {Probability-Voltage relationship based on MC simulation}
\label{Section:sbg:mcsimu}
\begin{figure}[tb!]
    \centering
    \includegraphics[scale=0.28]{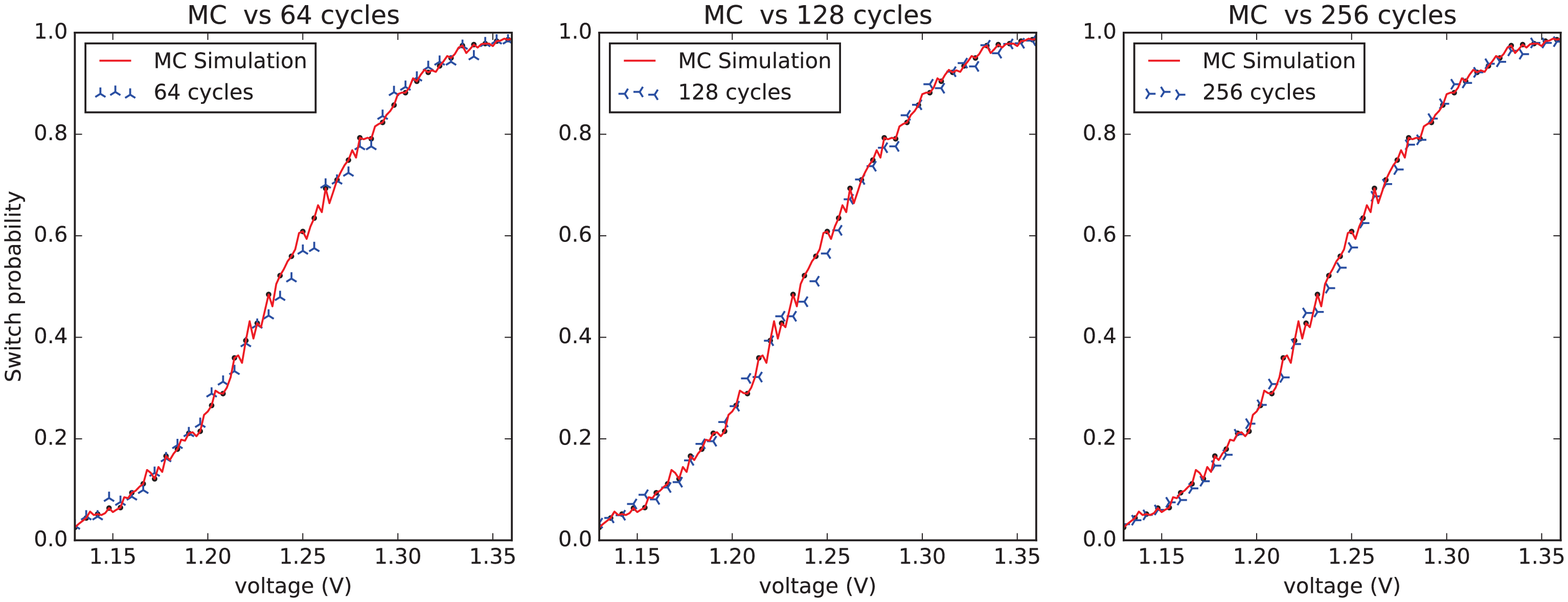}
    \caption{Applied voltage v.s. switching probability.}
    \label{fig:pvrelaton}
\end{figure}
SBG is used to generate SBs to represent probability number.
Different bias voltages correspond to different probability values.
In this section, the Probability-Voltage relationship of proposed SBG is analyzed using Monte-Carlo simulation strategy.
The simulation is processed by Cadence Virtuoso with 45 \textit{nm} CMOS and 40 \textit{nm} MTJ technologies.
In the simulation, a behavioral model of MTJ considering the stochastic switch feature is described by Verilog-A language \cite{wang2014compact}.
The write duration time is set to be 5 \textit{ns} because the relationship of voltage and probability is closed to linear under this setting.
The reset duration time is set to be 10 \textit{ns} in order to guarantee a 100\% reset switching.
For each bias voltage ranging from 1.13 \textit{V} to 1.36 \textit{V}, 1000 Monte-Carlo simulations are performed.
The simulated P-V relationship is illustrated in Fig.~\ref{fig:pvrelaton} by the red line.
From the figure we can find that as the increasing of voltage, the switching probability also increases monotonously.
It means that voltages and probability values are almost corresponding one by one.

\vspace{-4mm}
\subsection {Evaluation}
Two evaluation experiment results are presented in this section which prove that the stochastic bitstreams generated by the proposed SBG are high accuracy and low correlation.

\begin{figure}[tb!]
    \centering
    \includegraphics[scale=0.28]{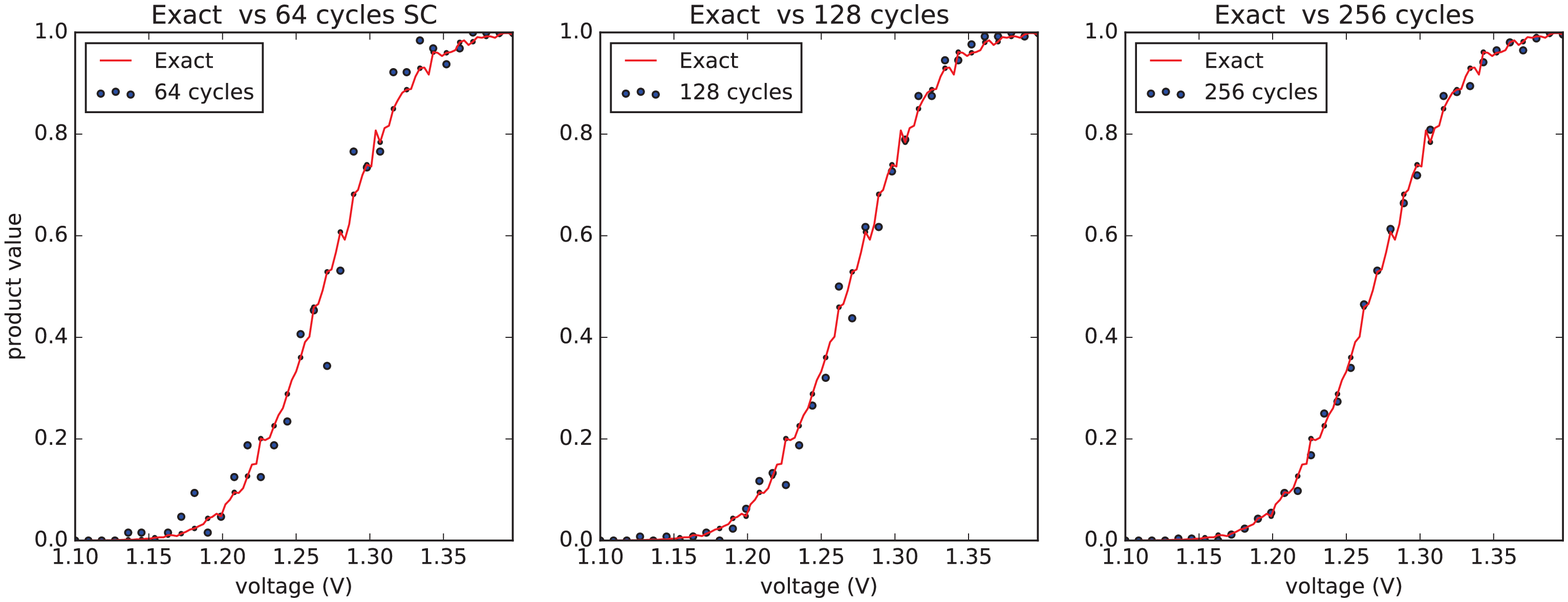}
    \caption{Comparison between arithmetic computing and stochastic computing.}
    \label{fig:product}
\end{figure}

Firstly, bitstreams are generated with length of 64, 128 and 256.
As shown in Fig.~\ref{fig:pvrelaton}, results of all the three classes bitstreams are well coincident with Monte-Carlo simulation results.
Compared with Monte-Carlo simulation results, the average errors are only 1.6\%, 1.3\% and 1.1\% for length of 64, 128 and 256, respectively.
It is obvious that the longer the bitstream, the smaller the error.
As described above, ``good'' bitstream requires low correlation with other bitstreams.
In the Verilog-A model, an effective seed generation strategy is integrated into MTJ model.
The strategy could guarantee that different MTJs use different seeds.
Because the seeds are independent of each other, there is no correlation between any two bitstreams.
To verify the random strategy, in the second experiment, a multiplication of two bitstreams driven by the same voltage is executed using AND gate.
Both the results of exact computing and stochastic computing with different bitstream lengths are shown in Fig.~\ref{fig:product}.
Statistical results show that the average errors are only about 2.8\%, 2.0\% and 1.2\%, respectively.

\label{Section:sbg:mcsimu}

So far, an SBG circuit is constructed based on MTJ and its efficiency has been proven by simulation results.
It is served as the most important component of the Bayesian inference system proposed in Section~\ref{Section:application}.

\vspace{-4mm}
\section {Applications}
\label{Section:application}
Different applications may be solved by different Bayesian inference mechanisms. Thus, structures of BIS are also different.
In this section, two different types of applications with different inference mechanisms are considered.
Using the MTJ based SBG and stochastic computing theory, we build two Bayesian inference systems for the two applications.

\vspace{-4mm}
\subsection {Data fusion for target location}
Data fusion is the process of integrating multiple data sources to produce more consistent, accurate, and useful information than that provided by any individual data source.
In this section, a simple data fusion example and corresponding Bayesian inference system are studied.

\subsubsection{Problem definition and Bayesian inference algorithm}
There are three noisy sensors on the $2D$ plane and each of them could provide two sensor data independently: Distance ($D$) and Bearing ($B$).
The problem is to calculate the object location $(x^\star,y^*)$ on the plane under the estimated data $(D_1, B_1, D_2, B_2, D_3, B_3)$.
The values of the problem parameters are similar to that in~\cite{coninx2016Bayesian} as following.
Three sensors locate at (0,0), (0, 32) and (32, 0) and the object actual position is (28,29).
For each sensor $i$, Given a position $(x, y)$, the distance model $p(D_i|x\,y)$ and bearing model $p(B_i|x\,y)$ satisfy the following Gaussian distributions:
\begin{align*}
p(D_i|x\,y) = N(\mu_{di}, \theta_{di}),  \quad
p(B_i|x\,y) = N(\mu_{bi}, \theta_{bi}).
\end{align*}
where, $\mu_{di}$ means the Euclidian distance between the sensor $i$ and position $(x,y)$, $\theta_{di} = 5 + \mu_{di}/10$.
And $\mu_{bi}$ is the viewing angle of the sensor $i$ and position $(x,y)$, $\theta_{bi}$ is set as 14.0626 degree.

The inference algorithm using sensor data can be expressed as $p(x\,y|D_1\,B_1\,D_2\,B_2\,D_3\,B_3)$.
Based on Bayes' theory,
\begin{equation}
\label{Equation:pro_simp}
\scriptsize
p(x\,y|D_1\,B_1\,D_2\,B_2\,D_3\,B_3)
\propto  p(x\,y ) * \prod_{i} { p(B_i|x\,y)p(D_i|x\,y) }
\end{equation}
In Eqn.~\eqref{Equation:pro_simp}, $p(x\,y)$ is known as the prior probability and the following six conditional probabilities are known as evidence or likelihood information. In this problem, the object may locate at any position. The prior probability $p(x\,y)$ has the same value for every position.
So $p(x\,y)$ is ignored in the following Bayesian inference system.

\subsubsection{Bayesian inference system}
\begin{figure}[tb!]
    \centering
    \hspace{-2mm}
    \includegraphics[scale=0.65]{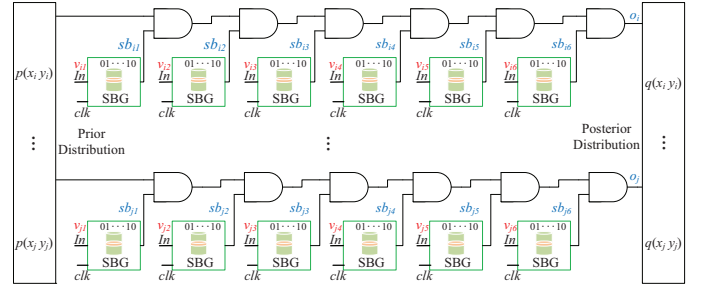}
    \caption{Bayesian inference system for object location problem.}
    \label{fig:df_system}
\end{figure}
It can be seen from Bayesian inference mechanism (Eqn.~\eqref{Equation:pro_simp}) that the distribution of object location is calculated by the product of a series of conditional probabilities.
In stochastic computing, this is processed using \texttt{AND} gates.
In addition, we could find that the calculation of probability value that the object locates at one position is independent for each other.
Based on the analysis, the Bayesian inference architecture is illustrated in Fig.~\ref{fig:df_system} as a matrix structure for this application.
For each position, $6$ SGBs are deployed to yield stochastic bitstreams and $5$ \texttt{AND} gates are deployed to achieve multiplication.
Thus, for a $64 \times 64$ grid,  $24576$ SBGs and $20480$ \texttt{AND} gates are needed.
In Fig.~\ref{fig:df_system}, the output of each row is the posterior probability value that the object locates at this position.
In our simulation, $64 \times 64$ counters are used to decode the outputs from stochastic bitstreams to float-point numbers by calculating the proportion of `1'.
The proposed system makes the best use of high parallel attribute of Bayesian inference and stochastic computing.
Utilizing the independent of inference algorithm (\textit{i.e.} Eqn.~\eqref{Equation:pro_simp}), all rows of the system could perform stochastic computing at the same time.
In each row, all the SBGs could yield bitstreams in parallel and the ``And'' operations are also implemented concurrently during reading the MTJ state.

\subsubsection{Simulation Results}
\begin{figure}[tb!]
    \centering
    \subfloat[Exact inference]{\includegraphics[scale=0.25]{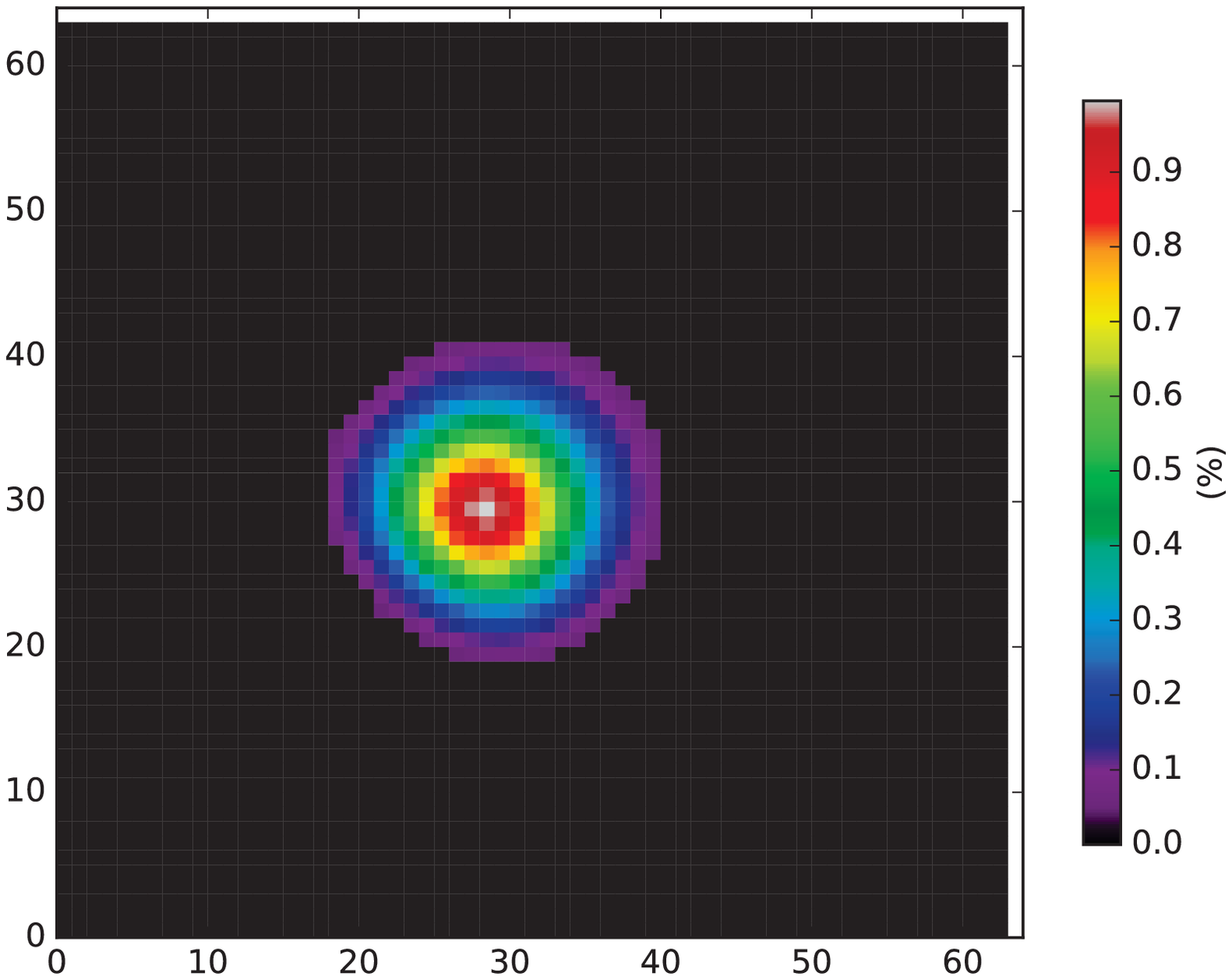}\label{fig:df:exact}}
    \hspace{+2mm}
    \subfloat[SB length = 64]{\includegraphics[scale=0.25]{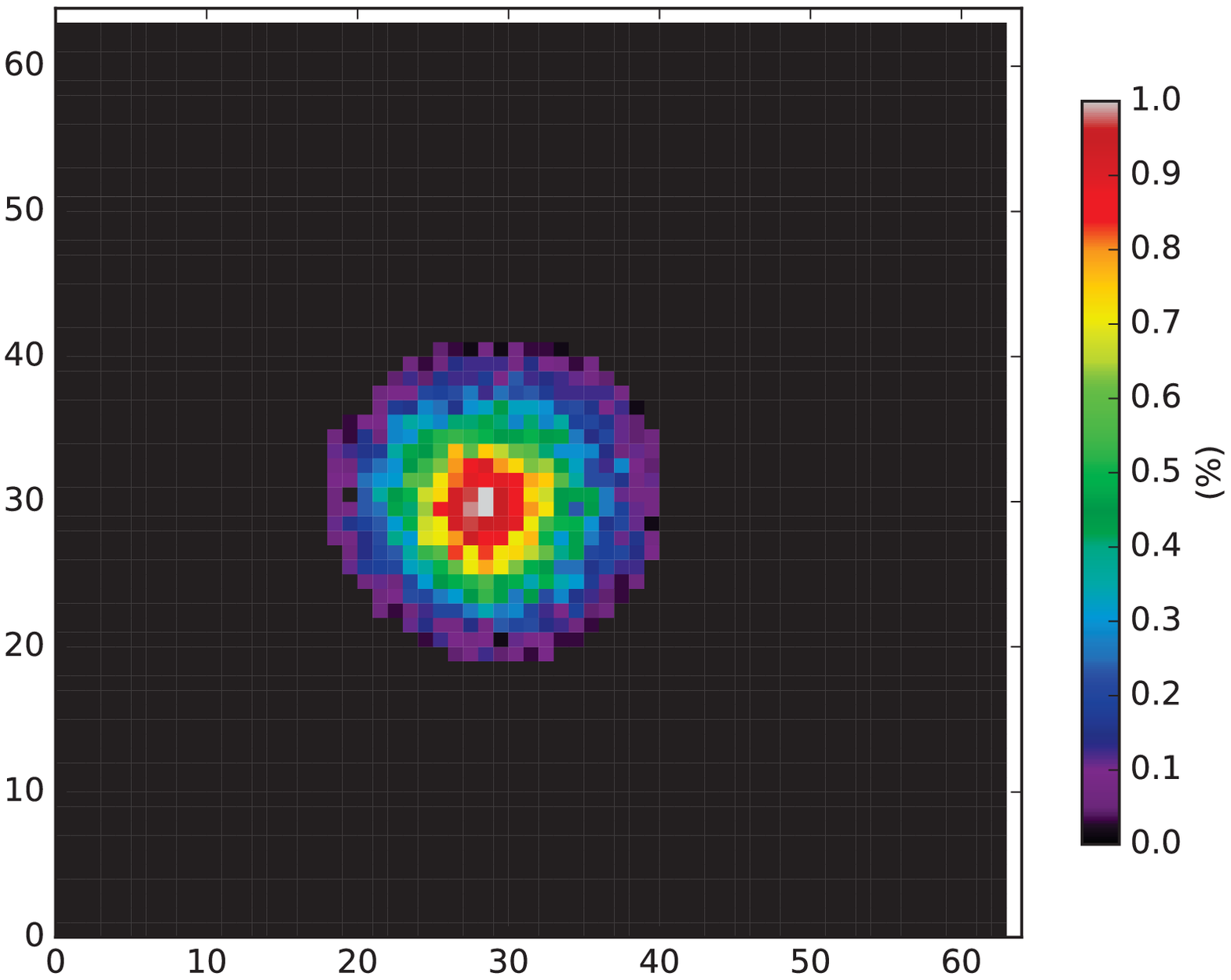}\label{fig:df:64T}} \\
	\subfloat[SB length = 128]{\includegraphics[scale=0.25]{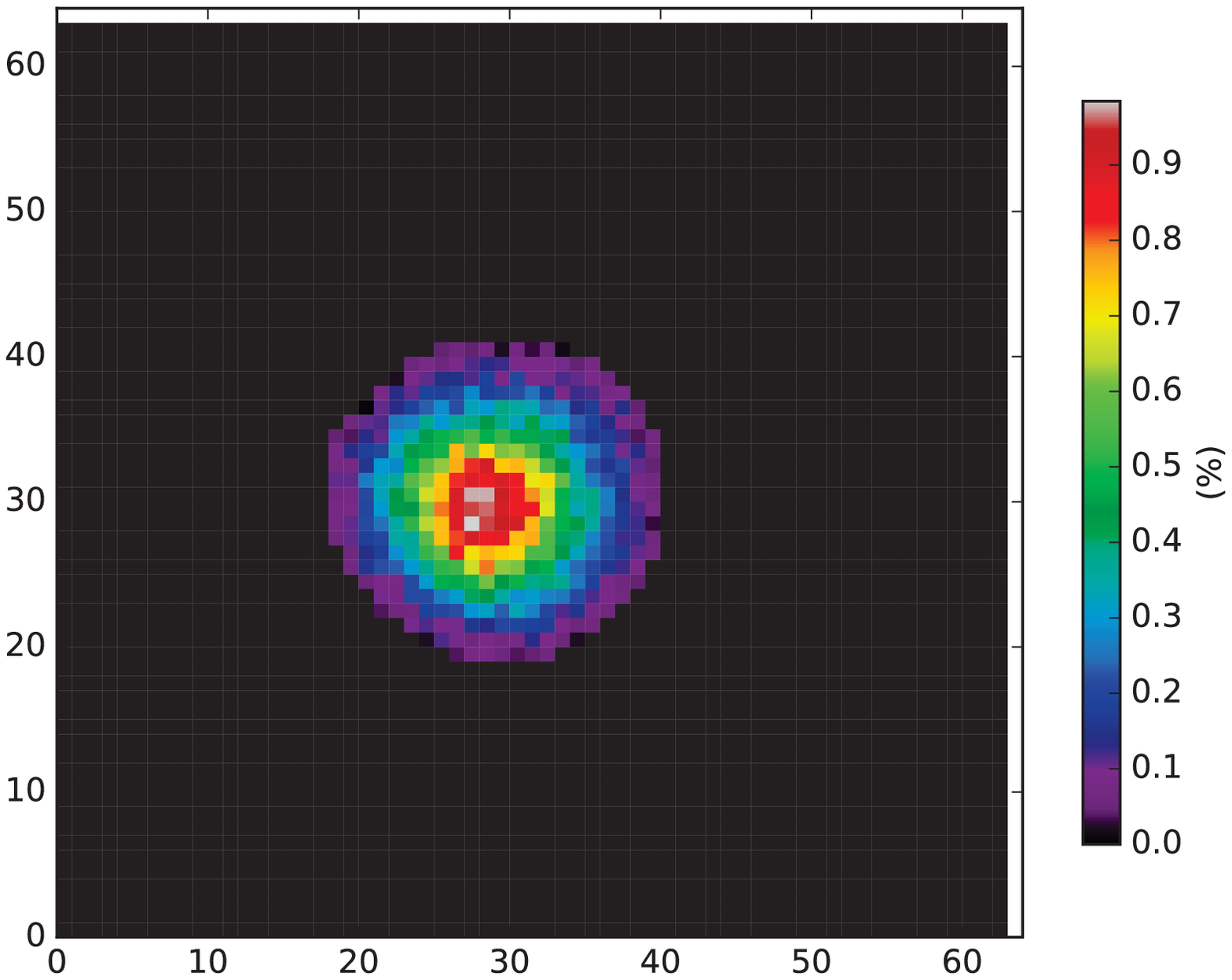}\label{fig:df:128T}}
	\hspace{+2mm}
	\subfloat[SB length = 256]{\includegraphics[scale=0.25]{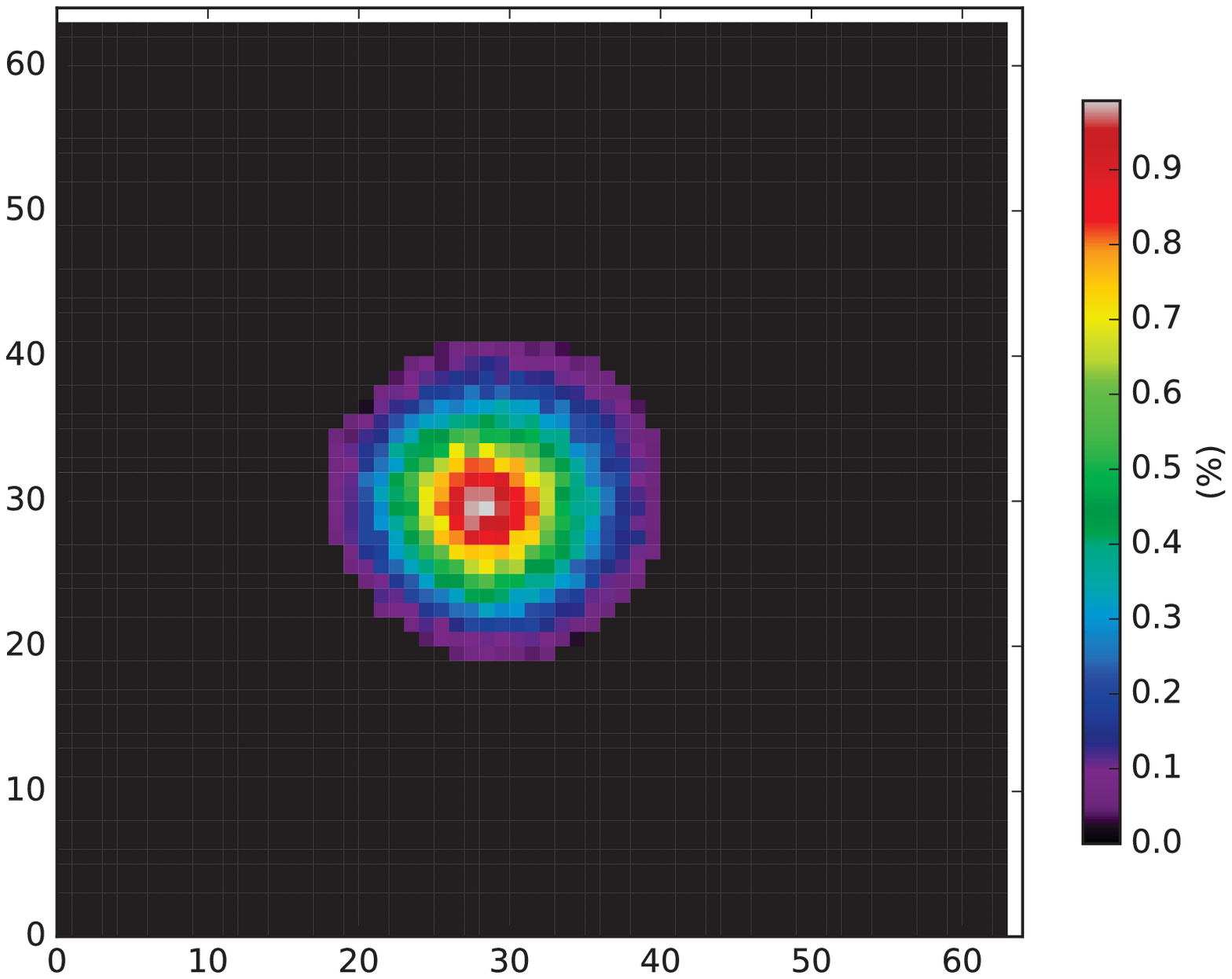}\label{fig:df:256T}}
    \caption{Data fusion result of target location problem on 64 $\times$ 64 grid. (a) Exact inference results. (b)-(d) Stochastic computing results with length of 64, 128, 256.}
    \label{fig:df_results}
\end{figure}

\begin{table}[tbp]
\centering
\caption{KL divergence analysis of target location problem}
\label{table:dfKL}
\begin{tabular}{cccc}
\toprule
\multirow{2}{*}{Grid Size} &              \multicolumn{ 3}{c}{Bitstream Length} \\
\cmidrule(lr){2-4}
 &         64 &        128 &        256 \\
\midrule
     64$\times$64 &    0.0090  &    0.0043  &    0.0018  \\

     32$\times$32 &    0.0086  &    0.0041  &    0.0019  \\

     16$\times$16 &    0.0080  &    0.0035  &    0.0011  \\
\bottomrule
\end{tabular}
\end{table}

Cadence Virtuoso is used to analyze the accuracy and efficiency of the proposed BIS.
In the simulation, $64 \times 64$, $32 \times 32$ and $16 \times 16$ grids are utilized to test our Bayesian inference system.
The finer the grid, the more accurate the target position.
For every grid scale, stochastic bitstreams with length of $64$, $128$ and $256$ are generated to perform stochastic computing.
The longer the stochastic bitstream, the higher the stochastic calculation accuracy.
In Fig.~\ref{fig:df_results}, four object location inferred results are shown by heat map on $64 \times 64$ grid.
Fig.~\subref*{fig:df:exact} is the exact inference result using arithmetic computing in float-point arithmetic computer.
Fig.~\subref*{fig:df:64T},~\subref*{fig:df:128T} and~\subref*{fig:df:256T} are the inference results by the proposed Bayesian inference system with stochastic bitstreams length of $64$, $128$ and $256$, respectively.
The simulation results indicate that the proposed system could achieve the Bayesian inference results correctly.
Compared with exact inference results, the longer the stochastic bitstream, the smaller the error.
To quantify the precision of the inference system, the Kullback-Leibler divergence (KL divergence) between stochastic inference distribution and the exact reference distribution is calculated.
As shown in Table~\ref{table:dfKL}, the first column shows the grid scale.
The following 3 columns are the KL divergence value for different bitstream lengths.
Taking $32 \times 32$ grid for example, $10^{-3}$ KL divergence requires length of 256.
But for the same precision, the work in~\cite{coninx2016Bayesian} requires length of $10^5$.
The outstanding results benefit from the high accuracy and low correlation bitstreams generated by the MTJ based SBG.
As reported in~\cite{coninx2016Bayesian}, for a problem with $32 \times 32$ grid, the software version on a typical laptop takes 919 mJ, and the FPGA based Bayesian machine only takes 0.23 mJ with stochastic bitstream length of 1000.
Benefiting from the low power consumption of MTJs and high quality of SBG, the proposed Bayesian inference system only spends less than 0.01 mJ to achieve the same accuracy with the 32 $\times$32 grid.
Speed of the proposed Bayesian inference system depends on the bitstream length.
Because of the high parallel, the whole inference process only takes 40T \textit{ns}, where `T' means the bitstream length.

\vspace{-2mm}
\subsection {Bayesian Belief Network}
Bayesian belief network is a probabilistic graphical model that represents a set of random variables and their conditional dependencies via a directed acyclic graph.
In this section, a Bayesian belief network for heart disaster is studied.

\subsubsection{Problem definition and Bayesian inference algorithm}
\begin{figure}[tb!]
    \centering
    \hspace{-2mm}
    \includegraphics[scale=0.8]{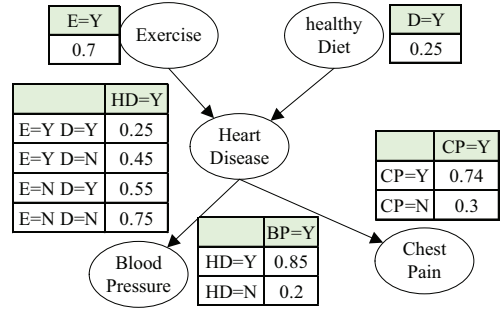}
    \caption{Bayesian belief network for heart disaster.}
    \label{fig:bbn_HD}
\end{figure}
Fig.~\ref{fig:bbn_HD} is a Bayesian belief network (BBN) example for heart disaster prediction.
In this network, the parent nodes of heart disaster (HD) are factors that cause heart disaster, including exercise (E) and diet (D).
The child nodes are clinical manifestations of HD, including blood pressure (BP) and chest pain (CP).
In addition to the graph structure, Conditional probability tables (CPT) are also given.
For example, the second value $0.45$ in the CPT of node HD means that if a person takes regular exercise but unhealthy diet, the risk of HD is $0.45$.
In this problem, we pay more attention to inference based on given evidences.
The inference mechanism could be classed as two groups based on the junction tree algorithm.
The first case is considering E, D and HD as a group and calculating $p(HD)$ as Eqn.~\eqref{Equation:p_cps}:
\begin{equation}
\scriptsize
\begin{aligned}
\label{Equation:p_cps}
p(HD=Y) = & [p(HD|E=Y D=Y)p(D=Y) + \\
& \; p(HD|E=Y D=N)p(D=N)]p(E=Y)+  \\
& [p(HD|E=N D=Y)p(D=Y) + \\
& \; p(HD|E=N D=N)p(D=N)]p(E=N)
\end{aligned}
\end{equation}
Here, Y means yes and N means No. If exercise or diet is determined, $p(E)$ or $p(D)$ in Eqn.~\eqref{Equation:p_cps} is 1, otherwise, is the value in CPT.
The second case is considering HD, HB and CP as a group and calculating $p(HD|HB\,CP)$ as Eqn.~\eqref{Equation:p_cxd}:
\begin{equation}
\scriptsize
\begin{aligned}
\label{Equation:p_cxd}
p(HD=Y|HB\,CP) = \frac {p(HB|HD=Y)p(CP|HD=Y)P(HD=Y)} {p({HB}\,CP)}
\end{aligned}
\end{equation}
The denominator of Eqn.~\eqref{Equation:p_cxd} can be calculated as Eqn.~\eqref{Equation:p_cxd_denominator}:
\begin{equation}
\scriptsize
\begin{aligned}
\label{Equation:p_cxd_denominator}
&p(HB|HD=Y)p(CP|HD=Y)P(HD=Y) + \\
&p(HB|HD=N)p(CP|HD=N)P(HD=N)
\end{aligned}
\end{equation}
Here, $p(HD=Y)$ is calculated by Eqn.~\eqref{Equation:p_cps}.
If HB or CP is diagnosed, the conditional probability value in Eqn.~\eqref{Equation:p_cxd} is the value in CPT, otherwise is 1.

\subsubsection{Bayesian inference system}
\begin{figure}[tb!]
    \centering
    \subfloat[]{\includegraphics[scale=0.8]{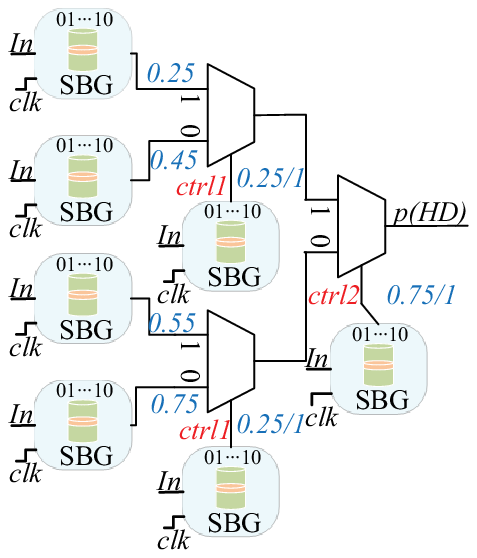}\label{fig:bbn_system:a}}
    \subfloat[]{\includegraphics[scale=0.8]{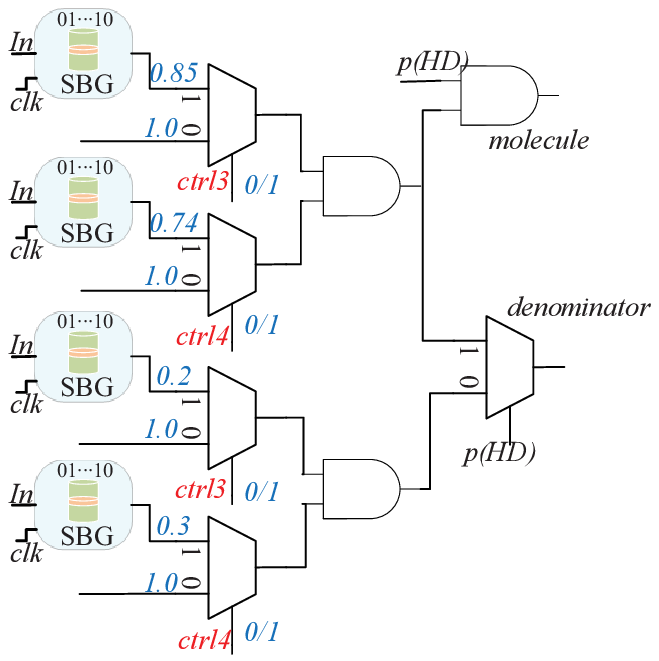}\label{fig:bbn_system:b}}
    \vspace{-2mm}
    \caption{Bayesian Inference system for BBN. (a) Implementation of Eqn.~\eqref{Equation:p_cps}. (b)Implementation of Eqn.~\eqref{Equation:p_cxd}.}
    \label{fig:bbn_archi}
\end{figure}
Based on the inference algorithm, the inference system could be easily constructed.
Eqn.~\eqref{Equation:p_cps} could be calculated by three \texttt{MUX}s as shown in Fig.~\subref*{fig:bbn_system:a}.
Eqn.~\eqref{Equation:p_cxd} could be calculated by three \texttt{AND} gates and five \texttt{MUX}s as shown in Fig.~\subref*{fig:bbn_system:b}.
Based on the evidence, the Bayesian inference is performed by different combination of \texttt{MUX} control signal.

\subsubsection{Simulation Results}
\begin{table}[tbp]
\centering
\caption{Example of BBN inference setting}
\label{table:bbn_test}
\begin{tabular}{lccc}
\toprule
Probability &      ($ctrl1$, $ctrl2$, $ctrl3$, $ctrl4$) &      \cite{ebay} &         SC \\
\midrule

  $p(HD|BP)$ &          (0.25, \, 0.75, \, 1.00, \, 0.00)    &     0.803 &   0.805 \\

 $p(HD|D,E,BP)$ &       (1.00, \, 1.00, \, 1.00, \, 0.00)    &     0.586 &   0.592 \\

 $p(HD|E,BP)$ &         (0.25, \, 1.00, \, 1.00, \, 0.00)    &     0.687 &   0.694 \\

 $p(HD|D,E,BP,CP)$ &    (1.00, \, 1.00, \, 1.00, \, 1.00)    &     0.777 &   0.742 \\

 $p(HD|CP)$ &           (0.25, \, 0.75, \, 0.00, \, 1.00)    &     0.703 &   0.700 \\

\bottomrule
\end{tabular}
\end{table}
The simulation of Bayesian inference system for BBN is also used Cadence Virtuoso and the simulation results are shown in Table~\ref{table:bbn_test}.
The first column of the table lists some the possible posterior probability.
The second columns gives the corresponding settings of control signal for each MUX.
Column 3 shows the exact results calculated by~\cite{ebay}.
Column 4 is the results calculated by the proposed bayesian inference system using stochastic computing.
The comparison between column 6 and column 7 indicates that the proposed Bayesian inference system for BBN could achieve reasonable results.

\vspace{-3mm}
\section{Conclusion}
\label{Section:Conclusion}
In this paper, a stochastic bitstream generator based on MTJ is proposed firstly.
Simulation results shows that the proposed SBG could yield ``good'' stochastic bitstreams.
Not only can the probability values be accurately expressed, but also the correlations between each other are low.
Based on MTJ based SBG and stochastic computing theory, two Bayesian inference systems for different applications are proposed.
Simulation results indicate that both the two systems could achieve high inference accuracy with fast running speed and low power consumption.
The future work will be carried on from two aspects.
The first one is further improving the performance of SBG in terms of accuracy, speed and power in order to build more efficient Bayesian inference system.
The second one is improving scalability to larger problems and widening extent of application.

\vspace{-3mm}


\end{document}